**Análise dos fatores determinantes para o decreto de prisão preventiva em casos envolvendo acusações por roubo, tráfico de drogas e furto.**
Um estudo no âmbito das cidades mais populosas do Paraná

**Giovane Cerezuela Policeno**[1]

**Mario Edson Passerino Fischer da Silva**[2]

**Vitor Pestana Ostrensky**[3]

**Resumo:** Partindo dos pressupostos teóricos do Labeling Approach, da Criminologia Crítica e da Economia Comportamental, levando-se em consideração que quase metade da população carcerária brasileira é composta por indivíduos que cumprem prisão preventiva, buscou-se avaliar quais as características dos flagranteados conduzidos às audiências de custódia apresentaram-se como determinantes para influenciar a subjetividade dos julgadores na decisão de determinar, ou não, a prisão preventiva desses. A pesquisa, inicialmente, adotou metodologia dedutiva, tendo como princípio que fatores objetivos externos estimulam os magistrados a decidir em determinado sentido. Focou-se então na identificação de quais características dos flagranteados e dos crimes, supostamente cometidos, seriam relevantes para nortear tais decisões. Posteriormente, deduzidos das decisões tais fatores, adotou-se uma metodologia indutiva, analisando os dados a partir dos pressupostos teóricos apontados. Durante a pesquisa foram analisadas 277 decisões, considerando decisão como decisão por indivíduo e não por audiência de custódia. Tal amostra embarcou decisões de seis juízes dentre as maiores cidades do Estado do Paraná e concernentes aos delitos de roubo, furto e tráfico de drogas. Concluiu-se então que a idade, gênero, classe social e o tipo de acusação que o flagranteado sofreu são determinantes para o decreto de sua prisão provisória, sendo que, a depender do juiz competente para o caso, as chances do decreto podem aumentar em até 700%, levando em consideração que as variáveis circunstanciais e causais sejam constantes. Diante da pequena amostragem, no que concerne ao número de juízes, fazem-se necessárias pesquisas mais amplas para que conclusões de validade nacional possam ser desenvolvidas.

**Palavras-chave:** 1. Prisão Preventiva; 2. Economia Comportamental; 3. Labeling Approach

*Abstract:* Based on the theoretical assumptions of the Labeling Approach, Critical Criminology and Behavioral Economics, taking into account that almost half of the Brazilian prison population is composed of individuals who are serving pre-trial detention, it was sought to assess the characteristics of the flagranteated were presented as determinants to influence the subjectivity of the judges in the decision to determine, or not, the custody of these. The research initially adopted a deductive methodology, based on the principle that external objective factors encourage magistrates to decide in a certain sense. It was then focused on the identification of which characteristics of the flagranteated and allegedly committed crimes would be relevant to guide such decisions. Subsequently, after deduction of such factors, an inductive methodology was adopted, analyzing the data from the theoretical assumptions pointed out. During the research, 277 decisions were analyzed, considering decision as individual decision and not custody hearing. This sample embarked decisions of six judges among the largest cities of the State of Paraná and concerning the crimes of theft, robbery and drug trafficking. It was then concluded that the age, gender, social class and type of accusation that the flagranteated suffered are decisive for the decree of his provisional arrest, being that, depending on the judge competent in the case, the chances of the decree can increase in Up to 700%, taking into account that the circumstantial and causal variables are constant. Given the small sample size, as far as the number of judges is concerned, more extensive research is needed so that conclusions of national validity can be developed.

---

1 Bacharel em Direito pelo Centro Universitário Curitiba. Pós graduando em Direito Penal e Criminologia pela UNINTER.
2 Acadêmico de direito do quinto ano da Universidade Federal do Paraná, facilitador em práticas circulares restaurativas, tendo matérias na área de direito penal e criminologia na Alma Mater Studiorum Università di Bologna.
3 Bacharel em Ciências Econômicas pela Universidade Federal do Paraná. Mestrando em Desenvolvimento Econômico pela mesma instituição.

**Keywords:** 1. Preventive prision; 2. Behavioral Economics; 3. Labelling Approach.

## 1 INTRODUÇÃO

Rotineiramente se faz alusão ao poder judiciário como "justiça", inclusive tecnicamente, ele é dividido em "Justiça" Comum" e "Justiça Especial". Tal confusão conceital culminou no nascimento de um sinônimo, e este reflete o imaginário popular acerca de como o judiciário exerce (ou deveria exercer) suas atribuições.

A crença em um julgamento "justo" é o alicerce mais importante que confere ao judiciário o imenso poder que ele maneja. Pressupõe-se que a justiça fora feita quando um juiz conduz e, por fim, julga um caso conforme os ditames legais.

A modernidade cuidou de avançar ainda mais o rumo da relativização da "justiça universal", porém, declaradamente, o direito não abandonou a missão da busca pela efetivação da justiça. Ainda são formulados princípios e normas os quais, amparadas pelo discurso político (essencial à manutenção da crença na infalibilidade das instituições), pretendem conter a contaminação das decisões judiciais por fatores alheios ao direito. A exemplo desses fatores pode-se citar (pré)conceitos próprios da mentalidade do julgador, ou mesmo as pressões populares e midiáticas.

Nessa senda, parece claro que o legislador se esforçou para enviar aos destinatários de poder a mensagem de que o judiciário representa uma instituição confiável, dificilmente influenciável (pois é independente[4]) e acima de manipulação (pois deve ser inerte, imparcial e fundamentar suas decisões conforme a lei[5]). Com relação à imparcialidade, novamente, são previstos, ainda, remédios jurídicos contra o julgamento de casos por juízes potencialmente parciais, como a "exceção de suspeição".

O princípio que talvez melhor ilustre o receio da contaminação política do judiciário é o Princípio do Juiz Natural. Embora não previsto expressamente na Constituição, tal princípio pode ser extraído do inciso LIII, do art. 5° e foi assimilado pela ordem jurídica, determinando inclusive a organização fixa dos Tribunais e âmbito da atuação de cada um (COUTINHO, 1998, p. 163). Em poucas palavras, o princípio apregoa que determinado caso deverá ser julgado pelo juiz competente predeterminado pela lei (COUTINHO, 1998, p. 163). Tratar-se de uma expressão da isonomia (COUTINHO, 1998, p. 163), visto que o critério de encaminhamento do caso de qualquer cidadão a determinado órgão jurisdicional se faz a partir da lei, possibilitando ao indivíduo saber quem o julgará antes de receber sua citação (conhecimento este pleno quando não há possibilidade de distribuição por sorteio).

Segundo Frederico Marques (2003, p. 197), os juízes seriam funcionários públicos *sui generis* cujo estatuto é a própria Constituição. É o juiz, palavra citada ao menos 500 vezes no Código de Processo Penal, o protagonista da relação processual, afinal, nas palavras de Marques (2003, p. 261), esse realiza a prestação jurisdicional ao cidadão. Jacinto Coutinho (2015), fazendo uso dos ensinamentos de Frederico Marques, apontou que o juiz é tão importante que ele não representa o Estado, mas o presenta, encarnando-o como órgão jurisdicional supra partes (além dos interesses destas) incumbente de efetivar os preceitos da ordem jurídica.

A doutrina contemporânea, contudo, concebe o juiz como um agente que reconstrói a realidade dos fatos, o qual deve ser imparcial, ao não ser influenciado pelo interesse de uma das partes, mas jamais poderá ser neutro (COUTINHO, 1998, p. 163). Assim o é, pois o juiz é um ser dotado de subjetividade, de ideologias pessoais e de experiências que invariavelmente norteiam seu modo de pensar e, se

---

[4] O artigo 2ª da Constituição da República Brasileira determina que o Poder Judiciário, assim como os demais, é independente.

[5] O art. 5°, inciso LXI, aponta como regra a necessidade da fundamentação, por parte de autoridade judiciária competente, como requisito essencial à validade de prisão fora do estado de flagrância. O artigo 93, inciso IX, estende a necessidade de fundamentação para todas as decisões dos julgadores e estipula, como regra, a publicidade dos julgamentos. O parágrafo único, inciso III, do art. 95, por sua vez, veda ao juiz a participação em atividades político-partidárias e o art. 99 garante autonomia administrativa e financeira ao judiciário.

o próprio direito é permeado de ideologias, não poderia ser diferente com seu operador mais destacado (COUTINHO, 1998, p. 163).

Uma das searas penais mais relevantes para se analisar criminologicamente os fatores "não jurídicos" que influenciam as decisões judiciais é a da homologação das decisões de Audiência de Custódia. Em tais audiências, o flagranteado (preso em flagrante) é conduzido à presença do juiz para ser entrevistado quanto às condições de sua prisão e, ao final, após manifestação de representante do Ministério Público e Defesa, o magistrado profere decisão acerca da legalidade, ou não do flagrante. Observa-se também que no *decisum*, ao entender a prisão como lícita, o julgador pode impor eventuais medidas cautelares e a principal delas é a prisão preventiva (responsável por 41% da população carcerária em 2016) (BRASIL; CNJ, 2016, p. 10).

É durante a audiência de Custódia, portanto, que o magistrado tem contato direto com preso, interroga-o sem a oitiva judicial de testemunhas, e via de regra[6], sem a existência de um processo.

Feitas as referidas ponderações, este breve ensaio não visa reiterar o óbvio, de que o juiz é também conduzido pelas próprias convicções ao canetear suas decisões. O objetivo fora avaliar quais as características dos flagranteados conduzidos às audiências de custódia e de seus crimes, supostamente cometidos, apresentavam-se como determinantes para influenciar a subjetividade dos julgadores na decisão de determinar, ou não, a prisão preventiva desses.

A pesquisa, inicialmente, adotou uma metodologia dedutiva, partindo-se do pressuposto de que fatores objetivos externos estimulam o juiz a decidir em determinado sentido, e assim focou-se na identificação de quais destes fatores (como características do flagranteado e da acusação do crime) seriam os mais relevantes para tanto.

Posteriormente, deduzidos das decisões tais fatores, adotou-se uma metodologia indutiva a partir da análise dos dados sob a ótica da Criminologia Crítica, do Labeling Approach e da Ciência Comportamental.

---

6 Os Centros de Audiência de Custódia estão recebendo também os casos de prisão efetivada por meio de mandado de prisão, de modo que, nestes casos, poderia existir ação penal anterior à prisão do acusado. No caso da presente pesquisa, foram analisados apenas casos de prisão em flagrante, os quais não teriam ação penal anterior à prisão.

## 2 HEURÍSTICA DA REPRESENTATIVIDADE

Há uma grande pressão popular no Brasil por uma justiça com mais celeridade e eficiência. Em resposta a isso, o Conselho Nacional de Justiça (CNJ) elabora metas anuais a serem cumpridas por cada instância do sistema judiciário. A título de exemplo, a Meta Nacional n°1 para o ano de 2017, incidente sobre todos os segmentos do judiciário, é, justamente, "Julgar mais Processos do que os Distribuídos (BRASIL, 2017). Nesses termos, com o aumento da cobrança institucional, os juízes são compelidos a julgar mais em menos tempo (critério meramente quantitativo).

Observa-se que os juízes assim decidem com base três fatores limitantes: i) informação, ii) capacidade de avaliação da mente humana e iii) tempo, variáveis essas que, segundo Simon (1955), explicam o motivo de, nas tomadas de decisões, a racionalidade ser limitada. Ou seja, não é possível julgar utilizando todas as informações, mesmo que essas estivessem disponíveis (e não estão).

Conforme Tversky e Kahneman (1975), em um processo decisório que envolve diversos elementos, o processamento de informação tende a ser complexo. Neste caso, para decidir rapidamente e cumprir o objetivo em questão, os sujeitos não raro recorrer a atalhos mentais, os quais são chamados de heurísticas.

Na maior parte do tempo esse fato torna o julgamento correto mais fácil e prático, porém as heurísticas geram ilusões cognitivas que podem afetar sistematicamente o julgamento. Ainda, se algum fator diferente daqueles legalmente relevantes for utilizado regularmente para a tomada de decisão do juiz, pode haver um viés sistemático no processo judicial.

Estudos ilustram as heurísticas e o viés que ocorrem no âmbito da tomada de decisões por parte dos juízes (GUTHRIE *et al.*, 2002; 2007) (PEER; GAMLIEL, 2013). Neste ensaio serão discutidos trabalhos específicos, que são o foco desta pesquisa.

O primeiro e mais relevante neste contexto é a heurística de representatividade, a partir da qual o indivíduo atribui uma representação a uma determinada categoria. Com a representação formada, um evento pode ser definido dentro da categoria por ter atributos semelhantes a essa, mesmo sem gozar da característica primordial que forma a tal categoria (KAHNEMAN;TVERSKY, 1972).

No caso desta pesquisa, essa heurística pode induzir à avaliação inconsciente de que uma pessoa está, ou não, dentro da representação de um criminoso. Os vieses gerados por esse "atalho" mental podem distanciar a decisão daquela que seria "técnica". Com isso, podem ser criados esteriótipos a determinados grupos de pessoas que tenham aspectos em comum, conforme descrito por Zaffaroni *et al.* (ZAFFARONI *et al*, 2003, p. 46):

> "Por tratar-se de pessoas "desvaloradas", é possível associar-lhes todas as cargas negativas existentes na sociedade sob a forma de preconceitos, o que resulta em fixar uma imagem pública do delinquente com componentes de classe social, étnicos, etários, de gênero e estéticos."

Isso não implica que o indivíduo não levará em consideração as demais informações disponíveis, mas, com a impressão já existindo, ele adotará outros fatores que corroborem com o pré-julgamento (BODENHAUSEN; WYER, 1985).

Há uma importante discussão sobre como o sistema penal de diferentes países ocidentais, em média, julga diversamente os cidadãos de acordo com suas etnias, punindo com maior probabilidade e rigor os afro-americanos e latinos. Geralmente os juízes negam agir deste modo, apesar de forte evidência estatística em sentido contrário, como em Bushway e Piehl (2001) e Abrams *et al* (2012). Considerando tanto a evidência empírica, quanto a veracidade do depoimentos dos magistrados, Lopez (2000) argumenta que os membros da instituição jurídica americana, por meio de processos mentais inconscientes, perpetuam o racismo em suas decisões.

Outro viés relevante levantado em pesquisas sobre o sistema judiciário é acerca do gênero do(a) acusado(a). Existem duas correntes divergentes sobre a diferença no rigor dos juízes para com as acusadas do gênero feminino. A primeira delas refere-se ao papel de gênero, apregoando que os juízes punem com maior rigor as mulheres por entender que elas teriam um papel de subordinação na sociedade perante o homem. Assim, uma maior punição serviria de exemplo para reforçar o comportamento feminino adequado (CHESNEY-LIND, 1977).

A teoria do Cavalheirismo/Paternalismo, pelo contrário, afirma que as mulheres são menos punidas, diante da concepção de que são frágeis e vítimas em potencial, merecendo, portanto, proteção (FRANKLIN; FEARN, 2008). Essa corrente pode ser interpretada como resultado do uso da heurística de representatividade – consciente, ou não - por parte dos juízes. Como a mulher não é, a princípio, a representação do típico criminoso, a tendência é que o uso dessa heurística a favoreça.

Recorrentemente também se discute acerca da influência da idade dos flagranteados no processo decisório dos magistrados. Doerner e Demuth (2010), entre outras conclusões, demonstraram que o jovens são mais suscetíveis a punições severas do que indivíduos mais velhos, o que é reforçado por outros autores, como Champion (1987), Cutshall e Adams (1983). Steffensmeier *et al* (1995), por sua vez, aponta uma relação curvilinear. Os mais propensos às punições seriam os acusados entre 21 e 29 anos, sendo que, a partir desta idade, a relação se tornaria negativa conforme os acusados fossem mais velhos, pois, inclusive, a representação padrão de alguém mais velho dificilmente é associada a um perfil criminoso. Já os mais jovens - entre 18 e 20 anos – gozariam de alto nível de leniência.

Também é possível a ocorrência de determinado viés devido à heurística da disponibilidade. Nela, o tomador de decisão analisa a probabilidade de ocorrência de um evento com base na facilidade com que ele consegue lembrar de um exemplo do mesmo. Os mesmos vieses citados acima podem ser derivados dessa heurística, já que os juízes tratam de vários casos, por dia, de pessoas com características semelhantes.

Outra heurística que potencialmente influecia nos casos em questão é a de ancoragem. Seu efeito ocorre quando o sujeito faz um ajustamento em sua decisão com base em um padrão pré determinado. Como a dificuldade de se afastar da âncora é grande, o indivíduo, em média, fará apenas um ajuste pequeno. Portanto, a decisão final é, inconscientemente, afetada pela âncora determinada (TVERSKY; KAHNEMAN, 1975).

Em uma audiência de custódia, o promotor, anteriormente à decisão do juiz, se manifesta a favor da prisão preventiva, ou da liberdade provisória. A manifestação do promotor pode acarretar em um efeito de ancoragem, como descrito por Peer e Gamliel (2013). Assim, há uma grande possibilidade de, dependendo da sugestão, a decisão do juiz ser diferente da que seria caso não o promotor não houve intervido.

Importante ressaltar que existe a possibilidade do uso de heurísticas para fatores que, teoricamente, não deveriam influenciar o julgamento. Algumas variáveis instrumentalmente relevantes de cada caso podem ser alvo dos atalhos. Por exemplo, o juiz pode sobrevalorizar o tráfico de uma droga específica no peso de sua decisão, seja por lembrar de casos na mídia, ter experiências familiares relativas a casos com drogas,por exemplo.

Não será o objetivo do artigo identificar as heurísticas possívelmente utilizadas, mas sim identificar os vieses existentes nas audiências de custódia. Porém, pela maior razoabilidade de interpretação, serão considerados pela heurística de representatividade.

## 3 REFLEXÕES CRIMINOLÓGICAS ACERCA DOS FATORES POTENCIALIZADORES DA PRISÃO PREVENTIVA

Conforme será apresentado posteriormente, várias homologações de prisão realizadas pelos juízes foram embasadas em fundamentos conceituais imprecisos, como a "garantia da ordem pública, ou econômica", conforme apregoa o art. 312 do Código de Processo Penal. Assim sendo, é interessante relembrar as ponderações de Christie (2011, p. 16), reflexo dos pressupostos das Criminologias Críticas, de que o crime, em si, inexiste, representando apenas uma das diversas formas de classificação, dependente do contexto cultural, de ações tidas como reprováveis, e que se trataria, portanto, de um conceito funcional de crucial utilidade para o exercício do controle social de condutas.

Zaffaroni e Pierangeli (2006, p. 60) completam as referidas ponderações ao afirmarem que toda conceituação é limitada, mas a verdade é infinita, e nesses termos, toda referência à verdade é necessariamente parcial, ou seja, ao se definir um ato como criminoso, seja na esfera legislativa ou judicial, a definição em si possui uma carga ideológica, o que é evidente, pois o próprio direito penal é cultural e valorativo (ZAFFARONI; PIERANGELI, 2006, p. 89). Ocorre que cada ideologia detêm o seu padrão ideal de ser humano e baseia suas concretizações na premissa de que tudo está justificado em virtude de uma necessidade, que implica na efetivação deste padrão (ZAFFARONI; PIERANGELI, 2006, p. 61).

A referida temática começou a ser trabalhada, no âmbito da criminologia, quando o objeto desta deixou de ser "quais são as causas do crime", até então tratado como ente ontológico, para "como explicar as causas de criminalização".

Retrocedendo um pouco, o início do desenvolvimento da ciência criminológica brasileira sofreu grande influência da Criminologia Positivista Italiana, o que acarretou, do final do século XIX até as primeiras décadas do XX, na concepção de que o crime correspondia a uma doença (RAUTER, 2003, p. 12). Nessa senda o perigosismo, que seria uma corrente ideológica latinoamericana proveniente da doutrina Criminológica Clínica de Nina Rodrgues, apresentava a pena como tratamento, e os principais doentes seriam os indivíduos provenientes de classes sociais menos abastadas (ZAFFARONI; PIERANGELI, 2006, p. 310).

Trata-se do perigo que os diferentes podem gerar à integridade daqueles que se atentam às leis, e, então, a manutenção do sistema penal cumpriria uma função substancialmente simbólica em face dos marginalizados doentes, sustentando simbolicamente uma estrutura de poder através da via punitiva (ZAFFARONI; PIERANGELI, 2006, p. 71). O sistema penal, então, reproduziria as injustiças e estratificações sociais (ZAFFARONI; PIERANGELI, 2006, p. 309).

Zaffaroni e Pierangeli (2006, p. 315) então apontam como função do direito penal a garantia de um ambiente externo seguro para a auto-realização humana no que tange ao que é considerado necessário para ser realizado em âmbito de convivência. Esse conceito implica de que o citado ramo do direito pretende proteger determinado modo de vida, o qual, para a criminologia crítica e o labeling approach, corresponde ao modo de vida de quem o formula e de quem opera.

Dando continuidade ao raciocínio, a vertente do labeling approach, ou enfoque/teoria do etiquetamento, trata especialmente do aspecto reacional em frente às condutas, trazendo a grande contribuição de que seria justamente a reação provocada em frente a um dado que acarretaria na sua qualificação como delito, ou não (PRADO; MAILLO, 2004, p. 319). Sob a ótica do interacionalismo simbólico, o qual dialoga com o labeling, a criminalização ocorreria quando determinado grupo inicia um processo de definição ao interpretar uma conduta como desviante, passando então a determinar que o sujeito praticante é um desviante e, por fim, aplicando ao indivíduo um tratamento específico (BARATTA, 2002, p. 94).

Tem-se então a construção do criminoso e da criminalidade, na ótica de Fritz Sack, como uma realidade social criada a partir de uma "qualidade" imposta por juízos atributivos (BARATTA, 2002, p. 107). Nessa linha, Sack aponta que o ato de condenar alguém perpetrado por um juiz, corresponderia à produção de uma nova realidade, pois a sentença tem o condão de redefinir a identidade dos réus ao imputar-lhes o *status* de "criminoso", por exemplo (BARATTA, 2002, p. 107).

Atualmente, o Brasil possui aproximadamente 13.790 magistrados no âmbito da "Justiça Comum" (os que não compõem os Tribunais Especiais) (O Globo, 2016). No Paraná a média salarial dos magistrados estaduais é de R$ 35.925,68. Tem-se então uma pequena parcela da população a qual goza de uma priveligiada posição econômica e social quando comparada com o restante da população brasileira.

Além do prestígio social, regado a pronomes de tratamento, auxílios financeiros e portentosos salários, os juízes contam com um poder imensurável, o qual, inclusive, os tornam uma ainda mais distinta: o poder de definição nos casos que lhes são apresentados.

Como já observado, o poder de definição pertenceria ao grupo social que promove o processo de definição, e no caso o grupo por excelência é o dos juízes, e o processo em questão pode também ser chamado de "Criminalização Secundária" (PRADO; MAILLO, 2004, p. 322). Enquanto as instâncias legislativas realizam a criminalização primária ao tipificar condutas, à polícia e ao judiciário cabe a criminalização secundária ao definirem quais dos comportamentos observados na realidade material corresponderiam à concretização de tipos penais, vulgo: práticas de crimes. Para Zaffaroni e Pierangeli (2006, p. 70), é justamente a criminalização secundária que colabora para a construção do estigma do delinquente, afinal, o deliquente é resultado da seletividade proveniente da citada crimalização, anteriormente este status não lhe havia sido conferido.

Como o labeling e o interacionismo simbólico limitam-se às descrições dos processos de crimalização, para se buscar informações acerca das relações de poder que conduziriam tais dinâmicas (BARATTA, 2002, p. 160), faz-se necessário recorrer a algumas vertentes da Criminologia Crítica.

A Criminologia Radical compreende que fundamento do direito penal é garantir a hegemonia de um grupo social (ZAFFARONI; PIERANGELI, 2006, p. 70), o qual deteria o poderio econômico e político. Parte-se da premissa então que o direito penal, quando é aplicado, pune com intensidade desigual, distribuindo o status criminoso de modo desigual (BARATTA, 2002, p. 162), tendendo à consolidação de um direito penal do autor (focado nas qualidades do réu tanto quanto, ou cima, das qualidades do ato).

Cristina Rauter (2003, p. 19) apresenta o aparelho judiciário como a instância, por excelência, de possibilidade de exploração de um grupo social por outro. Para a autora, seria no âmbito do judiciário que os magistrados, membros da elite, tratados como seres detentores da verdade absoluta, reproduziriam as segregações sociais a partir de uma racionalidade própria de uma classe mais abastada, a qual, diante da ainda vigente ideologia do perigosismo, teme aquilo que se recusa a compreender e, assim, culpabiliza. A instância judicial, via de regra, não é um lugar de reflexão, mas sim de prescrição, de determinação. Ao gozar de tamanho poder, o juiz se afasta ainda mais da realidade dos julgados, a empatia que ele poderia sentir ao decidir não raro pode ser mitigada pela sensação da superioridade, e, conforme aponta Christie (2003, p. 19), abstraí-se que as consequências do julgamento recairão sobre um ser humano e todos vinculados a ele.

Via de regra, os juízes brasileiros são homens, brancos, socialmente abastados, de origem social de classes ricas, ou médias, então, obviamente, eles tendem a se identificar com os indivíduos de seu meio social muito mais do que com o favelado, com o pobre, com o simples. Questiona-se aqui então se seria possível falar em justiça quando se está sendo julgado por um membro de uma tribo com costumes tão diversos e de uma realidade incompatível, quando não, antagônica a sua.

A própria classe social dos juízes estigmatiza e teme aqueles pertencentes às camadas sociais menos abastadas e seria ingênuo crer que isso não afeta, ainda que inconscientemente, os juízos de valor dos magistrados. Criados em um meio mais seguro, a concepção da sociedade que deve ser defendida, na mente dos juízes, talvez não seja composta pelos pobres curtidos e menos instruídos com os quais eles raramente tiveram contatos profundos durante suas vidas.

Nesses termos, a sociedade que o juiz se propõe a proteger, o ente superior o qual os homens integram, não é a sociedade do pobre, mas sim a sua, e, como adendam Zaffaroni e Pierangeli (2006, p. 86), o direito penal é mais autoritário quando, em maior grau, a sociedade confunde-se com Estado (leia-se plano institucional).

## 4 Metodologia

Durante a pesquisa foram analisadas 277 decisões, considerando decisão como decisão por indivíduo e não por audiência de custódia. Tal amostra embarca decisões de seis juízes dentre as maiores cidades do Estado do Paraná e concernentes aos delitos de roubo, furto e tráfico de drogas. Os autos foram consultados via Projudi e obtidos mediante livre pesquisa no site do Tribunal de Justiça do Estado do Paraná. Em nenhum dos casos, ressalta-se, houve o relaxamento da prisão diante de qualquer ilegalidade do flagrante.

Os critérios analisados nas decisões subdividem-se em quatro grupos: i) dados do processo, ii) dados do delito em questão, iii) dados do flagranteado e iv) dados da decisão.

No primeiro grupo, incluem-se o número dos autos, a comarca e o juiz que proferiu a decisão.

No segundo grupo, foram analisados o tipo penal e eventuais majorantes e qualificadoras considerados pelo magistrado, bem como possíveis peculiaridades do caso concreto, abordadas no relatório do julgador, que pudessem apresentar alguma relevância à decisão, tais como: quantidade de drogas, local em que ocorreu o delito, emprego de violência física, dentre outros.

No terceiro grupo, os dados pessoais do flagranteado incluem: condenações transitadas em julgado que, em relação à prisão em flagrante, configurariam reincidência no caso de nova condenação penal e os respectivos delitos (denominado por "reincidência");condenações transitadas em julgado que, em relação à prisão em flagrante, poderiam configurar maus antecedentes e os delitos correspondentes (denominadas por "maus antecedentes"); ações penais e inquéritos policiais em curso e respectivos delitos (denominadas por "passagens"); e dados pessoais do flagranteado (sexo, etnia, idade, emprego e vícios).

Os dados de reincidência, antecedentes e passagens foram obtidos pela análise de certidões do sistema Oráculo que estavam juntadas aos autos no momento da audiência, e os dados pessoais do flagranteado foram colhidos dos respectivos autos de interrogatório e vida pregressa, lavrados pela autoridade policial responsável pela prisão do flagranteado.

Por fim, no quarto grupo, foi identificada a decisão final como decretação de prisão preventiva, ou concessão de liberdade provisória, bem como seus fundamentos e justificativas. No caso das decisões de prisão preventiva, para fins desta pesquisa, considerou-se como fundamento as condições expressadas pelo artigo 312 do Código de Processo Penal (garantia da "ordem pública" ou econômica, conveniência da instrução criminal, asseguramento da aplicação da lei penal e descumprimento de obrigações impostas por força de outras cautelares). Observa-se que as justificativas seriam materializações dos fundamentos, sendo identificadas como principais: a credibilidade da justiça, reincidência, gravidade do *modus operandi*, evitar reiteração delitiva, descumprimento de cautelares, acautelar meio social e a suposta periculosidade do flagranteado.

Para identificar o impacto, tanto das variáveis criminais (tipo do crime, reincidência...), quanto dos flagranteados (idade, etnia...), a metodologia empregada foi o uso da regressão logística binária. Isto se deve à natureza da variável dependente, que é qualitativa, já que o juiz decide pela prisão preventiva ou pela liberdade provisória. Como são apenas duas opções, o modelo é binário. Decidiu-se pela prisão preventiva representar a ocorrência do evento. Desta forma, a variável resposta (Y) é igual a 1, em caso de ocorrência do evento, e 0, caso contrário.

O objetivo do modelo Logit é estabelecer a probabilidade da observação estar no grupo de ocorrência do evento de acordo com as variáveis dependentes. Assim, o modelo irá obter a probabilidade da decisão de acordo com os parâmetros estimados através de maxima verossimilhança com a seguinte função (King, 2008):

$$Logit(p_i) = \ln\left(\frac{p_i}{1-p_i}\right) = \beta_0 + \beta_1 x_1 + \ldots + \beta_k x_{k,i}$$

A interpretação dos parâmetros do modelo de regressão logística não é trivial. Os coeficientes mostrarão, de acordo com cada variável, a probabilidade de ocorrência do evento sobre a probabilidade de não-ocorrência. Para avaliar o impacto dos parâmetros na probabilidade do evento, estes devem ser transformados por antilogarítmos.

O quadro 1 apresenta as variáveis explicativas do modelo e seus respectivos valores, média e desvio padrão.

Quadro 1 – Variáveis independentes utilizadas no modelo

| Variáveis | Valores | Média | Desvio Padrão |
|---|---|---|---|
| **Juiz (1,2,3,4 e 5)** | Igual a 1 se a audiência foi julgada pelo respectivo juiz, caso contrário, 0; | - | - |
| **Roubo** | Igual a 1 se o crime for de roubo, caso contrário, 0; | 0.393 | 0.489 |
| **Sexo** | Caso o flagranteado seja do gênero masculino:1, se for feminino:0; | 0.866 | 0.340 |
| **Etnia** | Caso o flagranteado seja negro ou pardo:1, se for branco: 0; | 0.480 | 0.500 |
| **Idade** | Idade do flagranteado; | 25.953 | 7.429 |
| **Vício_maconha** | Se o flagranteado seja viciado em maconha:1, caso contrário, 0; | 0.209 | 0.407 |
| **Vício_cocaína** | Se o flagranteado seja viciado em cocaína:1, caso contrário, 0; | 0.0397 | 0.195 |
| **Vício_crack** | Se o flagranteado seja viciado em crack:1, | 0.173 | 0.379 |

| | caso contrário, 0; | | |
|---|---|---|---|
| **Reincidência** | Quantidade de reincidências no mesmo crime; | 0.683 | 1.227 |
| **Passagens** | Número de passagens pela polícia; | 1.169 | 1.473 |
| **Tráfico_maconha** | Caso o crime seja tráfico de maconha,1; caso contrário, 0; | 0.129 | 0.336 |
| **Tráfico_cocaína** | Caso o crime seja tráfico de cocaína,1; caso contrário, 0; | 0.086 | 0.281 |
| **Tráfico_crack** | Caso o crime seja tráfico de crack,1; caso contrário, 0; | 0.119 | 0.324 |
| **Residência** | Caso o furto ou roubo seja à residência da vítima,1; se não, 0; | 0.104 | 0.306 |

É importante explicar que durante a observação dos dados ficou nítida a diferença de probabilidade da decisão, em um sentido ou outro, entre os juízes. Por isso, para controlar esse fator, foi incluída uma variável *dummy* (variável que pode assumir valor 0 ou 1) para cada juiz.

Cabe ressaltar que como são analisados três tipos de crime na pesquisa (furto, roubo e tráfico) e eles são tratados como *dummies* na regressão logística, é preciso que um deles seja omitido para evitar a multicolineariedade[7]. Por isso, o furto é o "crime padrão", até por ser o de menor gravidade. O mesmo ocorreu para a variável de controle dos juízes, vez que um deles foi aleatoriamente omitido.

# 5 Resultados

---
7 Relações lineares exatas entre as variáveis dependentes.

Os resultados dos parâmetros estimados da regressão logística estão na Tabela 1. Também são apresentados o valor do teste de Wald, o qual demonstra a significância individual dos parâmetros, e as razões de chance (Odds ratio), que facilitam a interpretação dos coeficientes.

Tabela 1 – Resultados do modelo Logit: Impacto das variáveis na decisão do Juiz

| Coeficiente (N=277) | B | Wald | P | Razões de chance |
|---|---|---|---|---|
| Juiz 1 | -3.546*** | -5.13 | 0.000 | 0.028 |
| Juiz 2 | -1.742*** | -3.48 | 0.001 | 0.175 |
| Juiz 3 | -1.876*** | -3.19 | 0.001 | 0.153 |
| Juiz 4 | 1.999*** | 2.89 | 0.004 | 7.388 |
| Juiz 5 | -1.560* | -1.74 | 0.082 | 0.210 |
| Roubo | 2.557*** | 5.63 | 0.000 | 12.907 |
| Tráfico_Maconha | 1.270** | 2.25 | 0.025 | 3.561 |
| Tráfico_Cocana | 2.054*** | 2.82 | 0.005 | 7.805 |
| Tráfico_Crack | 1.457** | 2.40 | 0.017 | 4.293 |
| Sexo | 0.732* | 1.66 | 0.097 | 2.080 |
| Etnia | -0.118 | -0.35 | 0.727 | 0.888 |
| Idade | -0.055*** | -2.98 | 0.003 | 0.946 |
| Vício_Maconha | 0.691 | 1.53 | 0.125 | 1.99 |
| Vício_Cocaína | -1.800* | -1.71 | 0.088 | 0.165 |
| Vício_Crack | 0.054 | 0.11 | 0.911 | 1.056 |
| Reincidência | 0.751*** | 4.44 | 0.000 | 2.119 |
| Passagens | 0.244* | 1.95 | 0.051 | 1.277 |
| Residência | 1.179* | 1.87 | 0.062 | 3.251 |

Wald chi2(18) = 69.58   Prob > chi2 = 0.0000
Pseudo R2 = 0.3990
*significativo a 10%, **significativo a 5%, ***significativo a 1%.

O foco da análise deve se concentrar na direção dos parâmetros estimados, que indica se a variável aumenta ou diminui a chance do flagranteado ter a prisão preventiva decretada. A observação do valor absoluto dos coeficientes ficará em segundo plano.

O teste Wald global mostrou que o modelo tem significância conjunta. Já o pseudo $R^2$ expõe que quase 40% das decisões dos juízes é explicada pelas variáveis do modelo.

Como destacado anteriormente, fica evidente a diferença da chance de decreto de prisão preventiva entre os juízes. Por exemplo, caso o flagranteado seja julgado pelo juiz 4, ele tem 7.38 vezes mais chance de ser mantido preso do que o juiz utilizado como referência. Esse fator, utilizado aqui como variável de controle, tem potencial para ser melhor analisado em pesquisas posteriores.

As variáveis de controle do tipo de crime seguem o resultado esperado. O roubo tem maior probabilidade (12.9 vezes mais) de resultar em prisão preventiva do que o furto. O mesmo acontece com o tráfico. É importante notar que o tráfico pode ocorrer (e normalmente ocorre) com mais de uma substância ao mesmo tempo. Por isso não se pode afirmar que tem menor probabilidade de ser punido do que o roubo.

É razoável apontar que o tráfico de cocaína tem menor chance de prisão preventiva comparando com o de maconha e crack (2.19 e 1.81 vezes mais, respectivamente). As variáveis de reincidência, passagens e de crime em residência retornaram valores positivos, como esperado. Elas podem ser analisadas tanto do ponto de vista "legal", quanto do comportamental e serão melhor discutidas na próxima seção.

O resultado do coeficiente estimado do sexo do acusado mostra que os homens tem mais probabilidade de prisão preventiva do que as mulheres (2.08 vezes mais). Isso indica um viés sistemático por parte dos juízes e corrobora com a tese de cavalheirismo/paternalismo. A mulher é vista, segundo essa teoria, como alguém mais frágil que não está dentro da representação padrão de criminoso, o que pode revelar um provável recurso à heurística de representatividade.

As variáveis de etnia e de vício em maconha e crack não foram significantes, portanto não se diferencia os valores de 0.

Já a Idade retornou um coeficiente bastante relevante. Cada ano a mais na idade do flagranteado representa 5,4% menos chance de prisão preventiva. Isso vai de acordo com a representação padrão de um criminoso, em geral jovem, como encontrado por outras pesquisas, como em Doerner e Demuth (2010).

O vício em cocaína retornou a um valor negativo, ou seja, o flagranteado que alegou ter vício em cocaína tem mais chance (83,5%) de ser liberado provisóriamente. Essa variável pode ser interpretada, possívelmente, como uma *proxy* de renda, já que tal droga tem um preço elevado e, por isso, pessoas com melhor situação financeira dispõe de maior facilidade de obtenção da mesma. Assim, apesar de não haver possibilidade de confirmação, pode-se levantar a hipótese de que existe um viés de classe social nos julgamentos.

A Tabela 2 apresenta o grau de classificação correta pela estimação do modelo de regressão logística. Assim, utilizou-se um ponto de corte de 0,5. Ou seja, se o valor estimado for acima de 0,5 a previsão é de decisão por prisão preventiva. Deste modo, o modelo estimou de maneira acertada 82,67% das decisões.

Tabela 2 – Classificação preditiva do modelo Logit

| Estimação/Real | PP | LP | Total |
|---|---|---|---|
| **PP** | 145 | 27 | 172 |
| **LP** | 21 | 84 | 105 |
| **Total** | 166 | 111 | 277 |
| **Corretamente classificado** | | | 82.67% |

## 6 DISCUSSÃO

Nesta seção, pretende-se relatar e problematizar outras situações perceptíveis a partir da análise das decisões que, embora não sejam quantificáveis para os fins desta pesquisa, possam apresentar análises interessantes a partir da teoria comportamental e da criminologia, bem como **possíveis** justificativas para determinados resultados encontrados.

A primeira situação flagrante sobre a análise das decisões pode ser denominada como "massificação de modelos de decisão" ou, em outras palavras, o uso rotineiro e constante de modelos prontos de decisão utilizados por parte dos magistrados. Conforme brevemente comentado em seções anteriores, a carga de trabalho e a exigência de decisões rápidas impostas aos juízes supostamente criariam a necessidade do uso de modelos que já contenham o conteúdo de uma decisão. Dessa forma o magistrado já possui um modelo de decisão pronto mesmo antes de conhecer o caso concreto. Tal fenômeno dificulta o controle e análise do processo decisório, uma vez que a utilização maciça de modelos já contém argumentos prontos, como se fossem "coringas". Não se pretende imputar qualquer tipo de responsabilização aos magistrados, porém é inegável que a utilização dos modelos, não raro, leva à tomada decisões abstratas, o que, especialmente nos casos de prisão preventiva (que tratam da supressão da liberdade antes de sentença transitada em julgado) se mostra como uma técnica com grande potencial de lesar, injustamente, os jurisdicionados.

Outro aspecto perceptível na utilização dos modelos, na senda da hipótese ora tratada, sobretudo nas decisões que concedem liberdade provisória, é o aparente processo argumentativo apresentado. Na maioria dos casos em questão, as decisões de liberdade provisória costumam apontar, como argumento, a ausência de requisitos para a prisão, o que poderia significar que o magistrado busca, antes de tudo, razões e motivos para decretar a prisão preventiva. Isso contraria a lógica mais banal das medidas cautelares, qual seja, de que a prisão deveria ser exceção à regra.

Mais uma vez, importante lembrar que a possibilidade da massificação de modelos dificulta a compreensão do processo decisório e argumentativo, porém cumpre ressaltar que, dada a importância e a gravidade causada pelas decisões proferidas pelo Poder Judiciário, é fundamental refletir acerca desse fenômeno.

Outro fato a ser explorado, conforme citado em seção anterior, é a questão da utilização da reincidência e das "passagens" como argumentos para decretação da prisão preventiva, vez que tais circunstâncias indicariam uma propensão do autuado a delinquir novamente. Aqui cabe uma análise crítica sob a ótica da garantia fundamental da presunção de inocência. A existência de tais dados no histórico do flagranteado apenas possui o condão concreto de provar uma única coisa: de que o flagranteado praticou crimes e teve envolvimento em outros inquéritos policiais, mas jamais poderá indicar, a partir da ótica das garantias fundamentais, que tal indivíduo retornará a praticar delitos.

Embora nos referidos casos e na jurisprudência pátria[8] sejam aceitas relativizações da presunção de inocência, é inegável que tal lógica segue os pressupostos semelhantes aos "discursos perigosistas", que buscam, a partir da lógica de risco, impedir ocorrência de crimes antes que eles ocorram.

Há também a problematização acerca da insegurança jurídica, com base na alta "variação decisória" entre os juízes, de tal modo que a decisão imposta ao flagranteado está lançada à sorte de qual juiz presidirá sua audiência de custódia. Não se pretende afirmar que todos os magistrados devem pensar de forma semelhante, até porque isso seria impossível, dada a natureza humana. Por outro lado, é inconcebível que, conforme relatado em seção anterior, o flagranteado possa ter sua chance de sofrer prisão preventiva multiplicada por sete apenas em razão do juiz que julgará seu caso.

## CONCLUSÃO

Após as referidas ponderações fica clara a influência dos estereótipos sobre as decisões dos magistrados nos casos de prisão preventiva, o que é evidenciado pela heurística de representatividade. Pelo que foi constatado pela econometria, jovens e homens tem maior probabilidade de terem prisão preventiva decretada. Além disso, é interessante notar que enquanto o tráfico de cocaína é o que tem a maior probabilidade de encarceramento, os indivíduos viciados na mesma droga têm menos chance de prisão preventiva para qualquer crime. Pode-se especular que, como a cocaína é uma droga cara e apenas acessível às classes mais altas, a classe social do indivíduo seria um fator relevante para os julgamentos. Tem-se que a mentalidade do magistrado, ainda que existam parâmetros legais comuns, é um fator extremamente determinante para o decreto, ou não, da prisão preventiva, visto que, no caso, a depender do juiz competente, as chances do decreto podem aumentar em até 700%, levando em consideração que as variáveis circunstanciais e causais sejam constantes. Por fim, fazem-se necessárias novas pesquisas sobre o tema, que possam abranger mais localidades e amostras, encontrando, assim, padrões sistemáticos no sistema judiciário brasileiro como um todo.

---
8 A título de exemplo, pode-se citar: STF - RHC 122647 SP - Relator: Min. Roberto Barroso. Data de julgamento: 19 de agosto de 2014. Órgão julgador: Primeira turma. Publicação: DJe-178 DIVULG 12-09-2014 PUBLIC 15-09-2014 e STF: HC 122090 DF. Relator: Min. Luiz Fux. Data de julgamento: 10 de Junho de 2014. Órgão Julgador: Primeira turma. Publicação: DJe-125 DIVULG 27-06-2014 PUBLIC 01-07-2014.

**REFERÊNCIAS BIBLIOGRÁFICAS**